\documentclass[12pt]{iopart}
\usepackage{graphicx,amssymb}
\begin{document}
\title[Ultra-high-energy cosmic
rays in the AGASA--HiRes--Yakutsk cluster]{Magnetic deflections and
possible sources of the ultra-high-energy cosmic rays in the
AGASA--HiRes--Yakutsk cluster }

\author{S.\,V.\,Troitsky
}

\address{
Institute for Nuclear Research of the Russian Academy of
Sciences,\\
60th October Anniversary Prospect 7a, 117312, Moscow, Russia
}
\ead{st@ms2.inr.ac.ru}
\begin{abstract}
The cluster of ultra-high-energy cosmic rays observed by the AGASA, HiRes
and Yakutsk experiments is studied with respect to possible deflections of
particles in regular magnetic fields. Best-fit positions of a potential
source of these clustered particles are found, with account of the
errors in energy estimation, both in the frameworks of particular models
of the Galactic magnetic field and treating the direction and the amount
of deflection as free parameters.
The study suggests that an unknown regular component of either Galactic or
extragalactic magnetic field may dominate over modelled
components in the direction of the cluster. Possible sources of the cosmic
rays in that direction are considered.
\end{abstract}
\date{}

\pacno{98.70.Sa}

\maketitle

\section{Introduction}
Clustering of arrival directions of the cosmic rays with energies
exceeding $4\cdot 10^{19}$~eV, evidence to which was originally
found~\cite{AGASA1996} on the basis of the early AGASA data, was subject
to numerous tests and reanalises both with the updated AGASA event list
and with the event sets of other
experiments~\cite{AGASA1999,Uchichori,TT:clustering}. All statistically
significant positive results were obtained by making use of the sets of
cosmic rays which included the original list of Ref.~\cite{AGASA1996}, the
one used to formulate the conjecture --- without these 36 events, the
cosmic-ray samples were simply depleted statistically at these
high energies. Recently, the world data set was supplemented by the events
detected by the HiRes experiment in the stereoscopic mode. The arrival
directions were determined with unprecedented accuracy and, by themselves,
did not exhibit any significant clustering~\cite{HiRes}. The attempts to
explain this apparent discrepancy between AGASA and HiRes included both
the criticism of the previously reported values of the probability of
chance clustering~\cite{Finley} and analyses which demonstrated the
absence of significant discrepancy at
all~\cite{Semikoz,JapaneseClustering}.

A subsequent combined analysis of the HiRes and AGASA
data~\cite{HiResFarrar} did not find strong autocorrelation. However, it
has been found that one event observed by HiRes in the stereo mode falls
in the middle of the AGASA triplet. Though the energy of this event, as
reported by the HiRes collaboration, is just below the $4\cdot 10^{19}$~eV
threshold of the AGASA sample, the possible systematic energy shift
between the two experiments (see e.g. Ref.~\cite{Blasi}) brings it above
the threshold. Another HiRes-detected cosmic-ray event from a direction
close to the cluster~\cite{Farrar} has energy between $10^{19}$~eV and
$2.8\cdot 10^{19}$~eV which exact value is unpublished. The latter fact
makes it difficult to include the event in the analysis of deflections;
together with the lack of published data of other experiments at these
energies this forces us to postpone the consideration of this event until
more information become available.

Very recently, the Yakutsk collaboration reanalised their data with novel
methods and tools, which resulted in improved precision in the
determination of both the energies and the arrival
directions~\cite{newYakutsk}. Remarkably, this reanalysis revealed another
event whose arrival direction is very close to the cluster under
discussion. In the present study, we consider five cosmic-ray events: the
AGASA triplet, the Yakutsk event and the HiRes event with published
energy.

The HiRes-AGASA quadruplet was studied in Ref.~\cite{Farrar}, where the
best-fit position with account of {\it random} magnetic deflections was
found. It has been demonstrated there that in the frameworks of the model
of the Galactic magnetic field by G.~Medina Tanco, the best fit
for the {\it regular} magnetic deflection is zero. As it has been pointed
out in Ref.~\cite{Farrar}, a more detailed analysis should take into
account the errors in experimental determination of the particle's
energies and, since the Galactic magnetic field is poorly known
(especially in the cluster direction), should treat both the direction and
the amount of the {\it regular} deflection as free parameters.

In this paper, we perform this kind of analysis of the {\it regular}
magnetic deflections in the AGASA-HiRes-Yakutsk cluster
for various
assignments of the electric charges of the primary cosmic particles. Apart
from a blind model-independent two-parametric fit, several particular
models of the Galactic magnetic field are considered.
We discuss also the implications of the
existence of this tight cluster, present a list of potential astrophysical
sources and decline the possibility that all five particles resulted
from a single interaction or decay of an extreme-energetic particle which
took place in the vicinity of the Earth.

The rest of the paper is organised as follows. In Sec.~\ref{sec:CRays}, we
list measured coordinates and energies of the five events, estimate
the experimental errors and discuss possible systematic shifts in the
energy scales between the experiments. We proceed in
Sec.~\ref{sec:unconstrained} with simulations of the best-fit positions in
a simplified but unconstrained model where the direction of the shift and
the amount of the deflection {\em per unit energy} are kept free parameters,
however common for all five cosmic-ray events. The fits with regular
magnetic shifts are, for some choices of parameters, better than
those for neutral particles or for random deflections only. In
Sec.~\ref{GMF}, we consider various
models of the Galactic magnetic field.
We vary now the amount of
deflection per unit energy, while the directions of deflections (now
slightly different for different members of the cluster) are taken from
the models. The results of Sections~\ref{sec:unconstrained},~\ref{GMF} and
their implications for the origin of the cosmic-ray particles which form
the cluster are discussed in Sec.~\ref{sec:sources}; the conclusions are
briefly formulated in Sec.~\ref{sec:concl}.

\section{The cosmic-ray events forming the cluster}
\label{sec:CRays}
The region of the sky close to the Beta Ursa Major star attracts the
attention of cosmic-ray physicists since the announcement~\cite{AGASA1999}
that the third cosmic-ray particle with energy in exceed of $4\cdot
10^{19}$~eV, observed by the AGASA collaboration, arrived from that
direction which hosted already a cosmic-ray doublet~\cite{AGASA1996}. With
the addition of the HiRes~\cite{HiResFarrar} and Yakutsk~\cite{newYakutsk}
events, this becomes the hottest spot for the search of possible
astrophysical sources of the ultra-high-energy cosmic rays. Notably, even
without the two events from the set used to formulate the clustering
conjecture~\cite{AGASA1996}, it is still a triplet (and maybe a
quadruplet, if a lower-energy HiRes event is included). The significance
of this clustering signal will be analysed
elsewhere.

The coordinates and energies of the three AGASA events were published in
Ref.~\cite{AGASA1999}. The error bars in energy estimations by AGASA are
taken to be about $18\%$ (see Ref.~\cite{AGASA:energy} for a detailed
discussion). For the arrival directions, we use the energy-dependent
$68\%$ errors published in Ref.~\cite{AGASA1999}; note that they are more
conservative than those used in Ref.~\cite{Farrar}, obtained from a
private communication.

As it has been stated in Ref.~\cite{HiResFarrar}, the energy determination
error in the HiRes stereo dataset does not exceed $20\%$, and $68\%$ of
events are reconstructed within $0.6^\circ$ (which corresponds to the
two-dimensional Gaussian distribution with the standard deviation of
$0.4^\circ$). We use these values in our simulations.

The error bars for the energy determination of the Yakutsk events are
given in Ref.~\cite{newYakutsk} on the event-by-event basis. For the event
under consideration, $23\%$ are quoted. The arrival direction is
determined with the accuracy of $3^\circ$. Note that compared to the
originally published data~\cite{Yakutsk:yellowBook}, the arrival direction
was shifted by about one degree and the energy was increased by a factor
of $1.5$.

All three experiments have observed a significant number of events with
energies studied here; their fields of view are quite similar; and still,
the energy spectra measured by them exhibit systematic difference. This
may be attributed to possible systematic shifts in energy determination
(see, for instance, Ref.~\cite{Blasi}), which may be easily estimated from
the reported fluxes. To treat the events observed by different experiments
on a uniform basis, it is necessary to account for these systematic shifts,
so we correct the HiRes and Yakutsk energies by shifting them to the AGASA
scale. Of course, the absolute choice of the scale is not important for
our study which treats the amount of deflection per unit energy as a free
parameter. The shift is $25\%$ up for HiRes and $7\%$ down for Yakutsk.

\begin{figure}
\begin{indented}
\item[]
\includegraphics[width=112mm]{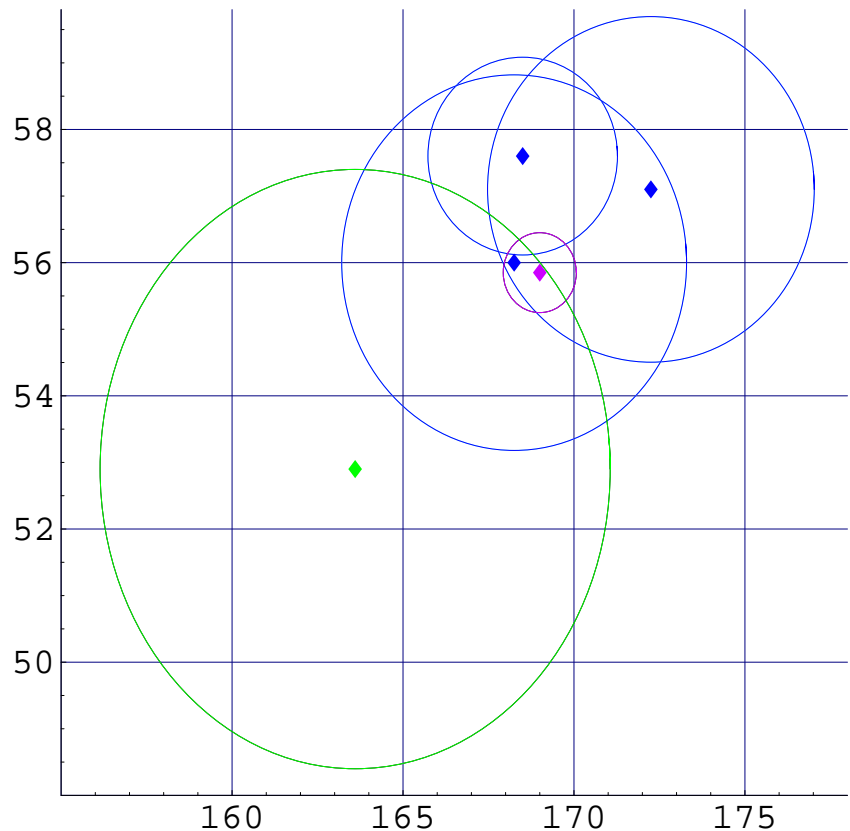}
\end{indented}
\caption{\label{Fig:cluster}The skymap of the cluster events with positional error circles.
The event with the smallest circle (maroon) is the HiRes one, the largest
circle corresponds to the Yakust event (shown in green), the other three
events (blue) were detected by AGASA. }

\end{figure}

The details of the five events are summarised in Table~\ref{Tab:events}
for reference.
\begin{table}
\caption{\label{Tab:events}Events forming the cluster. (1), the event code used in this
paper; (2), the name of the experiment; (3), the arrival date (dd/mm/yyyy);
(4), the energy reported by the experimental collaboration
(in units of $10^{19}$~eV); (5), the energy corrected to the AGASA scale
(in units of $10^{19}$~eV);  (6), the relative error in energy measurement;
(7) and (8), the arrival direction in the J2000 equatorial coordinates (in
degrees); (9), the accuracy of the arrival direction (in degrees).}
\lineup
\begin{indented}
\item[]\begin{tabular}{@{}ccccccccc}
\br
Event &Experiment & Date & $E$  & $E'$ &
$\Delta E/E$ & $\alpha^\circ$ & $\delta ^\circ$ &$\sigma ^\circ$\\
(1) & (2) & (3) & (4) & (5) & (6) & (7) & (8) &(9)\\
\mr
A1 &AGASA & 26/01/1995 & 7.76 & 7.76 & 0.18 & 168.5  & 57.6 & 1.0\\
A2 &AGASA & 01/08/1992 & 5.5  & 5.5  & 0.18 & 172.25 & 57.1 & 1.7\\
A3 &AGASA & 04/04/1998 & 5.35 & 5.35 & 0.18 & 168.25 & 56.0 & 1.9\\
H &HiRes & 29/04/2003 & 3.76 & 4.7  & 0.20 & 169.0  & 55.85& 0.4\\
Y &Yakutsk&17/03/1975 & 5.7  & 5.3  & 0.23 & 163.6  & 52.9 & 3.0\\
\br
\end{tabular}
\end{indented}
\end{table}
The arrival directions and their errors are plotted in
Fig.~\ref{Fig:cluster}.

\section{Best-fit positions of the source in the unconstrained
magnetic-field model}
\label{sec:unconstrained}
Deflections of charged particles in cosmic magnetic fields are very
important for studies of anysotropy in the distribution of arrival
directions and for searches of possible sources of cosmic rays.
However, these studies are limited by poor knowledge of cosmic magnetic
fields.

On its way from an extragalactic source to the detector, a charged
cosmic-ray particle is affected by the following components of the
magnetic fields. Firstly, there are (potentially very strong) magnetic
fields in the source -- these are clearly not important for the case of
distant sources seen as point-like ones. Then, there are intergalactic
magnetic fields; current theoretical and observational estimates of their
strength vary by orders of magnitude~\cite{IGMF:Sigl,IGMF:Dolag}. These
fields are random and result in a distortion of a ``cosmic-ray image'' of
the source. We note that the very existence of tight cosmic-ray clusters,
in assumption of a common origin of cluster members from a distant source,
favours the low intergalactic magnetic field~\cite{IGMF:Dolag} with
deflections not exceeding $0.1^\circ$ for the direction and energies of
interest. The dominant deflection is caused by the Galactic magnetic field
which, in turn, can be separated into the random and regular components.
Though the random component is locally stronger (see, e.g.,
Ref.~\cite{Beck2003} for a recent review), the deflection it causes is
most probably a subleading effect compared to that caused by the regular
component~\cite{TT:GMFrandom}.

The cluster of cosmic rays under consideration (without the Yakutsk event)
has been recently studied for the deflections caused by {\em random}
(intergalactic and Galactic) magnetic fields in Ref.~\cite{Farrar}. Since
the inclusion of the fifth event cannot change the conclusions
significantly, we restrict our attention to the {\em regular} magnetic
fields here.

As we will see in Sec.~\ref{GMF}, the latters are poorly known, especially
for the high Galactic latitudes where the cluster arrived from. There are
many particular models on the market, and the parameter estimates vary
widely. That is why we postpone the discussion of specific models to
Sec.~\ref{GMF} and perform here a blind model-independent test.

We assume that the deflections of the cosmic-ray particles in the cluster
are driven by two parameters: the direction and the amount of deflection
per unit energy, with values of parameters equal for all five particles.
This is a simplification since the comparison with particular GMF models
demonstrates (Sec.~\ref{GMF}) that the distance between the cluster
members is sufficient for these parameters to vary slightly (by a few
percent) from one arrival direction to another. However, this
parametrization is the only model-independent way to study the regular
deflections.

The deflection is inverse proportional to the particle's energy. As a
result, relatively large experimental errors in energy determination
affect strongly the amount of deflection and should be taken into account.

We assume that all five particles originate from a single source and do not
study the statistical significance of this claim. Under this assumption,
the best tool to search for the most probable source position is the
chi-square method which we adopt to incorporate the errors in energy
determination, not only in the determination of the arrival direction.

Let us denote by $\phi $ the position angle of the direction to which a
positively-charged particle is shifted, measured in the equatorial
coordinates clockwise from East from $0^\circ$ to $360^\circ$. The amount
of the deflection of a $10^{19}$~eV proton is denoted by $\epsilon $
degrees; that is the shift of a particle of energy $E$ and charge $q$ is
given by
\begin{equation}
\rho ={q\, \epsilon\, \over E_{19}}
\label{epsilon}
\end{equation}
 (hereafter, $E_{19}=E/(10^{19}~\rm{eV}) $).
To determine the probability for each particular position on the sky to
host the source of the cluster with {\em some} magnetic deflection, we
proceed as follows. We fix the coordinates of the position, $\alpha $ and
$\delta $; then we scan over $0^\circ \le \phi <360^\circ$ and $
0\le \epsilon <50^\circ$ to try different configurations of the magnetic
field. For each configuration, the arrival directions of cosmic rays are
shifted in the direction $\phi $ by the amount $\rho $ given by
Eq.(\ref{epsilon}) for the energy of a particle $E$ (corrected energy,
$E'$, for the HiRes and Yakutsk events). For each cosmic ray, a local
coordinate system is introduced which allows one to separate the angular
distance between the current point ($\alpha ,\delta $) and the corrected
arrival direction into two components --- $\Delta_{||}$, measured in the
direction of the shift, and the orthogonal $\Delta_\perp$ (both along the
great circles of the Celestial sphere). The chi square value is then a sum
over all cosmic rays,
$$
\chi ^2=\sum\limits_i \chi^2_i,
$$
 of the quantity
$$
\chi^2_i={\Delta_{i,||}^2 \over \sigma_i ^2 + \sigma_i ^{(\pm) 2}} +
{\Delta_{i,\perp}^2 \over \sigma_i ^2},
$$
where $\sigma _i$ is the standard error in the determination of the
$i$th arrival direction (given in Table~\ref{Tab:events}) and
$$
\sigma_i ^{(\pm)}={q \epsilon \over E_{19\,i} }\, {\Delta E_i \over E_i\pm
\Delta E_i}
$$
are determined from the conditions
$$
\rho (E_i \mp \Delta E_i)=\rho (E_i)\pm \sigma_i ^{(\pm)}.
$$
$\Delta E_i$ is the error in energy measurement determined from column (6)
of Table~\ref{Tab:events}. The choice of $\sigma ^{(+)}$ or $\sigma
^{(-)}$ is determined by the angle $\psi $ between the direction of the
shift and the direction from the shifted position to the current point
$(\alpha ,\delta )$: for $\psi <90^\circ$, we take the larger error
$\sigma ^{(-)}$, for $90^\circ \le \psi \le 180^\circ$ --- the smaller
error $\sigma ^{(+)}$. The discontinuity of $\chi ^2$ at $\psi =90^\circ$
does not arise because $\Delta_{||}=0$ there.

For each particular direction $(\alpha ,\delta )$, we minimise the $\chi
^2$ with respect to $\epsilon $ and $\phi $. If the errors were Gaussian,
the proper measure of the quality of the fit -- the probability of
this value of $\chi ^2$ to be obtained randomly -- would be given by a
chi-square distribution with 6 degrees of freedom (10 coordinates of the 5
events minus 4 parameters $\alpha $, $\delta $, $\epsilon $, $\phi
$)\footnote{We treat both coordinates as independent random variables with
$\sigma $ properly adjusted.
When
we assume zero charges, we have two parameters less and hence 8 degrees of
freedom. When one is  interested in the fit for a given location (e.g. a
candidate source), $\alpha $ and $\delta $ are fixed, and the number of
degrees of freedom increases by two. \label{footnote} }. In practice, the
errors are not Gaussian, but this probability $P$ is still an instructive
measure of comparative quality of fits, so we quote it in our results.
Better fits correspond to lower $P$ in these notations.

We considered the sets both with and without the Yakutsk
event. The latter fits are slightly better because of the relatively low
precision of the Yakutsk measurement, but both cases are consistent with
each other.
To test the stability of the results, we dropped also one of the AGASA
events - the A2, which arrived from a bit apart of the cluster center. The
results did not change considerably, though the quality of fits became
worse.

In Table~\ref{Tab:fits}, we present the best-fit positions of the sources
for
different charge assignments, motivated, in
particular, by the following considerations. The experimetnal data at the
energies of interest converge to the proton-dominated composition of
cosmic rays, so the most natural choice of the charge is $+1$. The presence
of the clustered component~\cite{AGASA:clustered-component} and
correlations of distant BL Lacs with HiRes stereo events at very small
angular separations~\cite{BL:HiRes} suggest a fraction of neutral
particles, hence we allow for zero charges. Finally, one may note that the
events, though they have very close energies, may be {\em aligned} along
some direction (more or less parallel to the Galactic plane in our case),
precisely like may happen in two particular models: in the Z-burst
model~\cite{Z-burst}, where protons, anti-protons and neutrons are equally
possible (to test this conjecture, we assign charges $-1$, 0 and $+1$ for
the corresponding primaries), and in the model~\cite{AGASA:nuclei}, where
the cosmic-ray flux is dominated by the remnants of photodesintegrated
nuclei, so that neutrons, protons and alpha-particles can be detected
(hence the allowed charges are 0, $+1$ and $+2$). We have thus chosen for
this study five most natural charge assignments which are listed in
Table~\ref{Tab:fits}. Other assignments are less motivated; some of them
were considered but resulted in worse fits.

\begin{table}
\caption{\label{Tab:fits}Best fits for the source positions with different charge
assignments. Columns: (1), the reference code of the charge assignment;
(2), charge assignment (see Table~\ref{Tab:events} for the event codes);
(3) and (4), the best-fit position in the J2000 equatorial
coordinates (in degrees); (5), the best-fit value of $\chi ^2$ and
the number of degrees of freedom; (6), the probability to obtain this
value of $\chi ^2$ randomly, assuming the chi-square distribution; (7) and
(8), the best-fit parameters of the magnetic deflection (in degrees).}
\lineup
\begin{indented}
\item[]\begin{tabular}{@{}cccccccccccc}
\br
Fit &\multicolumn{5}{c}{charges (2)} & $\alpha $ & $\delta $ & $\chi
^2$/d.o.f. &$P$ &$\epsilon $ & $\phi $\\
(1) & A1 & A2 & A3 & H & Y &(3) &(4)&(5)&(6)&(7)&(8)\\
\mr
0 & 0 & 0& 0& 0& 0& 169.0 & 56.1 & 6.44/8 & 0.40 & - & - \\
P & 1 & 1& 1& 1& 1& 167.6 & 62.7 & 2.83/6  & 0.17 & 35 & 265 \\
Z & 0 & 1& 0& 0&$-1$& 168.8 & 56.0 & 3.30/6  & 0.23 & 15 & 144 \\
N1 & 1 & 2& 1& 1& 0& 164.4 & 56.2 & 4.62/6  & 0.41 & 13 & 185 \\
N2 & 1 & 0& 1& 1& 2& 171.6 & 58.3 & 2.16/6  & 0.10 & 14 & 301 \\
\br
\end{tabular}
\end{indented}
\end{table}

\begin{figure}
\begin{indented}
\item[]
\includegraphics[width=102mm]{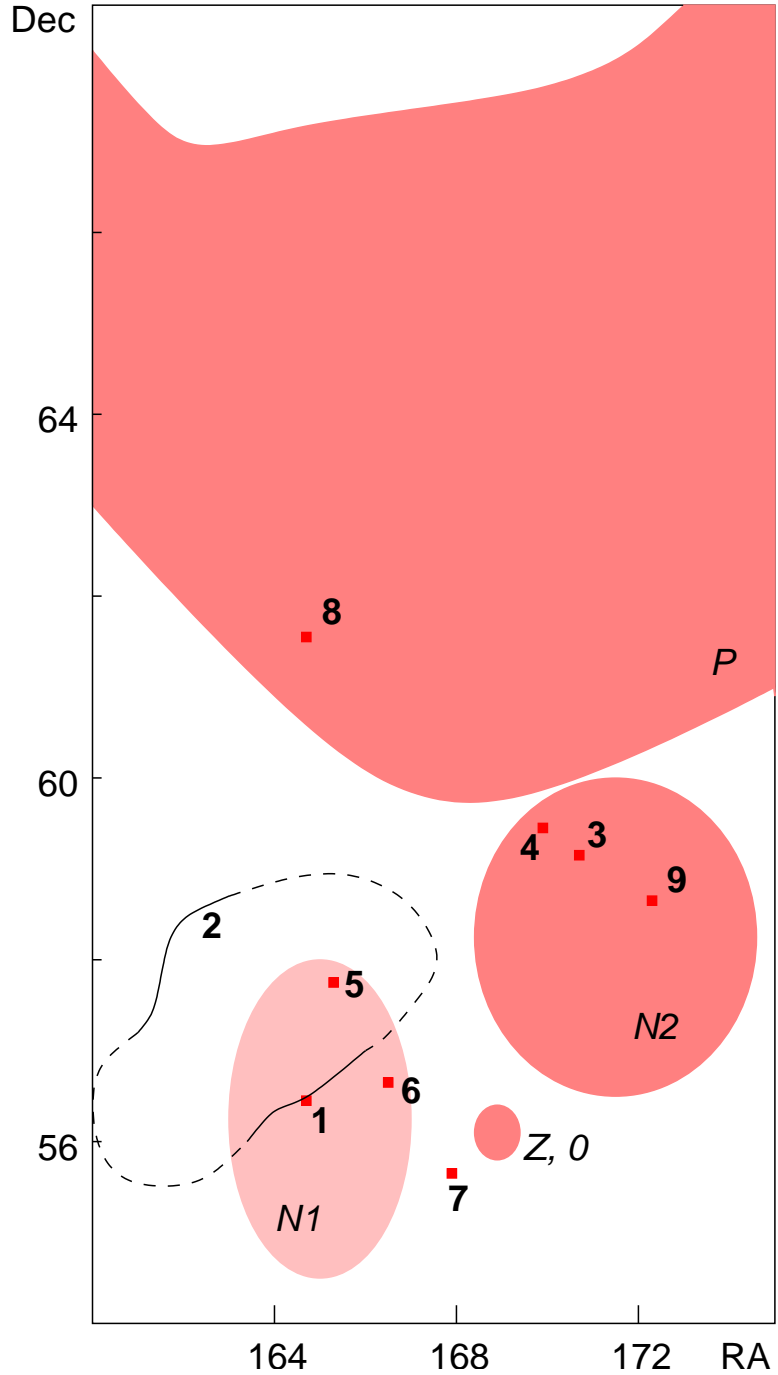}
\end{indented}
\caption{\label{fig:skymap}The skymap of the cluster surroundings. The shaded areas are the
regions of $1-P>68\%$ for different charge
assignments, indicated in italic (for the {\it 0} and {\it N1} fits, the
68\% areas are void; the 50\% area is presented by the light shading for
 the {\it N1} fit, while the similar area for the {\it 0} fit is located
 inside the 68\% area of the {\it Z} fit). The small boxes represent
 potential sources of the clustered cosmic-ray particles listed in
 Table~\ref{Tab:sources} (see Column 1 of Table~\ref{Tab:sources} for the
source labelling numbers, given in bold face). For the source 2, we plot
 the contour which includes the source with 99\% statistical
 probability~\cite{3EG} (the dashed part corresponds to extrapolation of
 the published contour). No potential sources were found outside the plot
 boundary for the extended area corresponding to the fit {\it P}.}

\end{figure}

The conclusion one may obtain from a look at Table~\ref{Tab:fits} is that
the inclusion of the magnetic field makes the fits better compared to the
case of neutral particles. Still, the precision is relatively low and the
fits are not overwhelmingly good. In particular, for good fits, the $90\%$
probability contours extend for several degrees which makes it difficult
to locate the possible source (see Sec.~\ref{sec:sources} for a discussion
and Fig.~\ref{fig:skymap} for the probability contour plots).

\section{Particular models of the Galactic magnetic field}
\label{GMF}
The cluster is located at high Galactic latitude, $b\approx 56^\circ$.
This implies that the Galactic magnetic field in that direction is
(a)~weak and (b)~poorly known. Indeed, the GMF models are
based dominantly on the measurements of the Faraday rotation of pulsars'
emission, and most of the pulsars are located in the Galactic plane, at
low latitudes (see e.g.\ a skymap in Fig.~8 of Ref.~\cite{Han1999}). The
impact of the field in the Milky Way disk is relatively small for that
large $b$ while the experimental information about the halo field is
currently far from being sufficient.

Let us summarize briefly contemporary GMF models, following the
reviews in Refs.~\cite{Beck2003,TT:GMF,HanI,HanII,ProuzaSmida,Vallee}.

The regular magnetic field in the disk follows the spiral structure of the
Milky Way. The models of the disk field
currently on the market are characterized by a few continuous parameters
determined by the local measurements with relatively good accuracy
(see Table~\ref{Tab:GMF}; we use the values of Ref.~\cite{TT:GMF}
consistent with the most recent data) and are classified into a few
discrete classes depending on the assumed global structure. The {\em
axisymmetric} (ASS) models assume similar field directions in different
spiral arms, while the {\em bisymmetric} (BSS) models assume field
reversals between the arms. Though current measurements favour the BSS
structure (see e.g.\ Ref.~\cite{Beck2003}), Ref.~\cite{Vallee} announced
their possible incorporation in an ASS type model to be formulated. We
consider models of both classes here. In the dipole-type (D) models,
the field direction in the Southern Galactic hemisphere is opposite to
that in the Northern one, while in the quadrupole-type (Q) models it is
the same.

The halo field is much worse known. Among commonly used approximations for
the field above and below the Galactic plane are the disk field multiplied
by a falling exponent~\cite{Beck2003,TT:GMF} with a scale height $h$ or by
a two-scale piecewise exponent~\cite{Stanev} parametrised by the scale
heights $h_1$ and $h_2$ and glued at the height $h_0$. They may represent
a part of the global halo field with a toroidal or poloidal structure,
strong near the Galactic Center (see Ref.~\cite{ProuzaSmida} for a
detailed discussion).

Recent studies of the magnetic deflections of ultra-high-energy cosmic
rays made use of the BSS field configuration with both Q and
D behaviour and a single vertical exponent~\cite{TT:GMF}, or with Q
behaviour and a two-step vertical exponent~\cite{Alvarez}. Recent
simulations~\cite{ProuzaSmida} included also the toroidal and poloidal
fields as well as some turbulent components. Here, we use both the
ASS and BSS models with both one- and two-step exponents. Q/D distinction
is irrelevant for the current study because the cluster is located in the
Northern Galactic hemisphere. For the toroidal field, we use the explicit
equations of Ref.~\cite{ProuzaSmida} for the maximal value of 1~mkG. We
approximate the poloidal field by exact magnetic-dipole field strength
normalized to have a local vertical component $B_z=$0.2~mkG and cut off at
1~kpc from the Galactic Center.

\begin{table}
\caption{\label{Tab:GMF}Parameters of the regular Galactic magnetic field. The second
column gives the experimental constraints on the parameters, the third
one gives the values adopted in this study. Note that $B_0$ is less than
the total local magnetic field strength due to presence of a random
component~\cite{Beck2003}.}
\lineup
\begin{flushright}
{\footnotesize
\begin{tabular}{@{}lcc}
\br
parameter & data & value used\\
\mr
\multicolumn{3}{c}{disk field:}\\
\mr
spiral structure & ASS or BSS & ASS or BSS\\
North-South reversal & D or Q & irrelevant\\
\mr
local strength of the {\em regular} field, $B_0$ &
\begin{tabular}{l}
$(1.4\pm 0.2)$~mkG~\cite{Beck2003}
\\
$(4\pm 1)$~mkG~\cite{ProuzaSmida}
\end{tabular}
&free parameter $B_0$\\
pitch angle, $p$ & $-8^\circ \pm 1^\circ$~\cite{Beck2003}&$-8^\circ$
\\
distance to the nearest field reversal, $d$ & $-0.2\dots
-0.6$~kpc~\cite{TT:GMF}&$-0.5$~kpc
\\
\mr
\multicolumn{3}{c}{field outside the disk:}\\
\mr
fall-off of the disk field& 1 or 2 exponents & 1 or 2 exponents\\
one-exponent height, $h$ & 1.5~kpc~\cite{Beck2003} & 1.5~kpc
\\
two-exponent heights
                    $\left\{
                            \begin{tabular}{l}
			     first, $h_1$\\
			     second, $h_2$\\
			     gluing, $h_0$\\
			    \end{tabular}
		     \right.$
&
\begin{tabular}{l}
 1~kpc~\cite{Stanev}
\\
 4~kpc~\cite{Stanev}
\\
 0.5~kpc~\cite{Stanev}
\\
\end{tabular}
&
\begin{tabular}{l}
 1~kpc\\
 4~kpc\\
 0.5~kpc\\
\end{tabular}
\\
proper halo field &toroidal, poloidal? & no or toroidal$+$poloidal\\
maximal strength of the toroidal field & 1~mkG~\cite{ProuzaSmida}
	  &$B_0/1.5$\\
          local vertical component of the field & 0.2~mkG~\cite{HanII} &
	  0.2$\cdot B_0/1.5$\\
\mr
maximal extent of the field, $R_{\rm max}$ & & 20~kpc\\
\br
\end{tabular}
 }
\end{flushright}
\end{table}

The procedure we implement for the analisys of particular GMF models is
very similar to that described in Sec.~\ref{sec:unconstrained}. Now, we
fix the direction of the shift according to a specific model (the
directions are allowed to be different for different members of the
cluster), but allow the amount of the shift (or, in other words, the
overall scale of the field strength) to be a free parameter. This approach
is justified by the fact that it is difficult to extract the strength of
the local {\em regular} magnetic field from the measurements, and
the experimental uncertainty in this parameter is very large (see
Table~\ref{Tab:GMF} for examples). The chi-square method allows one to
determine the best fit for the field strength in each particular model and
for a given charge assignment.
The results are almost insensitive to the model of the disk field (ASS or
BSS, one or two exponents) -- the chi-square values differ by a few per
cent. Hence, we quote (see Table~\ref{Tab:fitsGMFmodels}) only the numbers
for the most popular BSS model with one exponent, both with and without
inclusion of the less known toroidal and poloidal halo fields.

\begin{table}
\caption{\label{Tab:fitsGMFmodels}Best fits for the source positions in two models of the Galactic
magnetic field with different charge assignments. Columns: (1), the
reference code of the charge assignment; (2.1) and (3.1),
the best-fit values of $\chi ^2$ and the number of degrees of freedom;
(2.2) and (3.2), the probability to obtain this value of $\chi ^2$ randomly,
assuming the chi-square distribution; (2.3) and (3.3), the best-fit
local strength of the regular magnetic field (in mkG; minus sign indicates
the direction opposite to that given by a model). Columns (2.1)--(2.3)
and (3.1)--(3.3) refer to the BSS model with one exponent,
respectively without and with the toroidal and poloidal halo fields.}
\lineup
\begin{indented}
\item[]\begin{tabular}{@{}ccccccc}
\br
Charge &\multicolumn{3}{c}{without halo field}&
        \multicolumn{3}{c}{ with halo field}\\
assignment& $\chi ^2/$d.o.f. & $P$ & $B_0$
          & $\chi ^2/$d.o.f. & $P$ & $B_0$\\
(1)     &  (2.1)      &  (2.2)      & (2.3)
&  (3.1)      &  (3.2)      & (3.3)             \\
\mr
0 &  6.44/8 & 0.40 & -       &  6.44/8 & 0.40 & -       \\
P &  5.49/7 & 0.40 & $-0.98$ &  6.32/7 & 0.50 & $-0.15$ \\
Z &  6.05/7 & 0.47 & $-0.38$ &  5.11/7 & 0.35 & $-0.53$ \\
N1&  5.01/7 & 0.34 & $-0.60$ &  5.45/7 & 0.39 & $-0.23$ \\
N2&  6.44/7 & 0.51 & 0.      &  6.44/7 & 0.51 & 0.      \\
\br
\end{tabular}
\end{indented}
\end{table}

\begin{figure}
\begin{indented}
\item[]
\includegraphics[width=82mm]{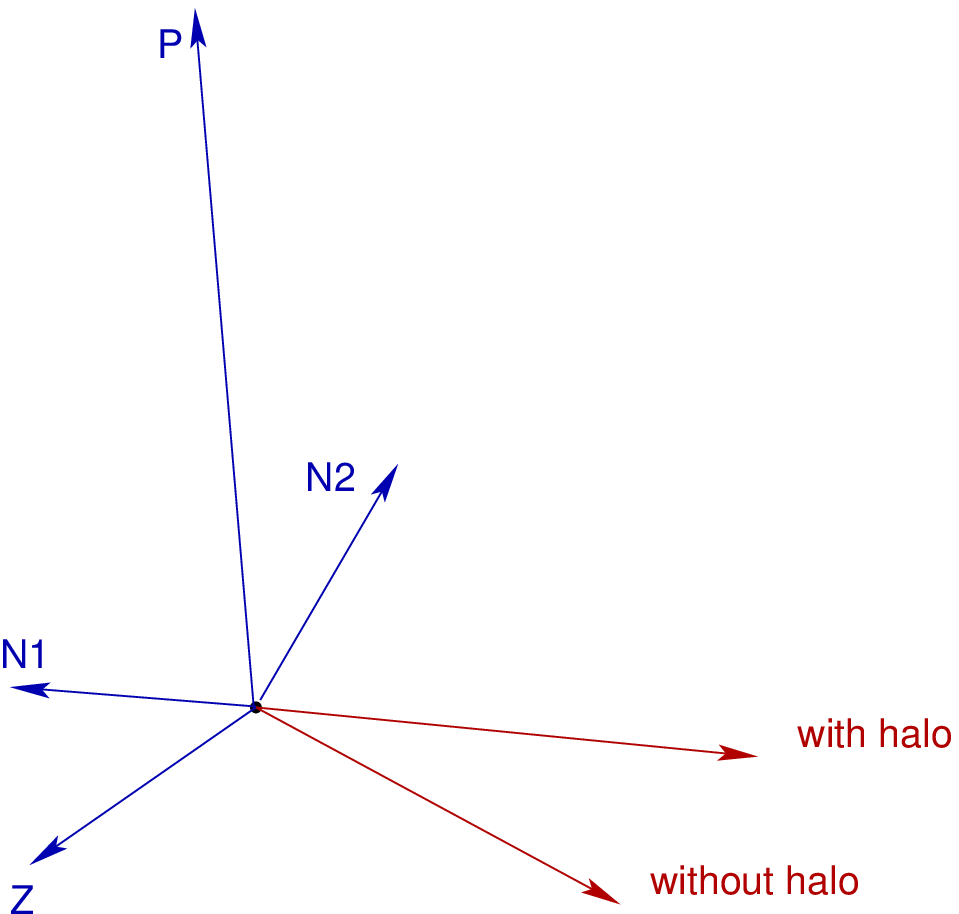}
\end{indented}
\caption{\label{fig:vectors}
Amounts and deflections of the shifts calculated for the best-fit values
for various charge assignments (P, Z, N1, N2) and for the models of the
Galactic magnetic field with and without inclusion of toroidal and
poloidal halo fields. The deflections for the GMF models were calculated
for the location of the H event assuming the BSS model with one-exponent
fall-off; for other models the results are very close.
The length of an arrow is proportional to $\epsilon $; the direction is
 given by $\phi $ (shift to the North, $\phi =270^\circ$, corresponds to
 the upward direction). }
\end{figure}

In the frameworks of all GMF models currently published, one
cannot achieve a fit for charged particles significantly better than the
fit for neutral ones (fit 0), in accordance with the results of
Ref.~\cite{Farrar} for a particular model. On the other hand, we have seen
in Sec.~\ref{sec:unconstrained} that \textit{ some} regular deflections
can make the fit much better. To illustrate this fact, we plot in
 Fig.~\ref{fig:vectors}
the directions and amounts of the shifts, parametrized by $\epsilon $ and
$\phi $, both for the best fits of Sec.\ref{sec:unconstrained} and for
GMF models considered in this section. We see that our study
may suggest that an unknown regular component of either Galactic or
extragalactic magnetic field is present.
This observation agrees with the conclusions of Refs.~\cite{TT:withoutGMF}
where a question was addressed, in which direction the cosmic rays are
shifted from their potential sources in different parts of the sky.
It would be interesting to perform
a careful study of the astrophysical data which may reveal the new
component independently.

\section{Potential sources and discussion}
\label{sec:sources}
Various potential astrophysical sources of ultra-high-energy cosmic rays
have been suggested in literature (see Ref.~\cite{Ancho:review} for a
recent review and Ref.~\cite{comparative} for a comparative study of
correlations between the cosmic-ray arrival directions and source
locations). In this section, we consider all possible sources around the
cluster direction which were listed in at least one of the catalogs of
Ref.~\cite{comparative}. We note that in this way, we restrict ourselves
to persistent extragalactic sources only.
There are nine
candidates in a large area spanned by the cluster position shifted by
magnetic fields with various charge assignments. They are listed in
Table~\ref{Tab:sources}  and plotted on a skymap, Fig.~\ref{fig:skymap},
 together with the best-fit positions of the cluster source for different
charges.

\begin{table}
\caption{\label{Tab:sources}Candidate sources around the cluster direction. (1), reference
number for the skymap; (2), object name; (3), object class (BLL$=$BL
Lac type object, Sy$=$nearby Seyfert galaxy, coll.$=$a pair of
colliding galaxies, $\gamma $-s.$=$ a gamma-ray source, IR$=$a luminous
infrared galaxy);
(4), the probability for the class (3) to correlate with the
 AGASA $E>4\cdot 10^{19}$~eV cosmic rays by chance, calculated on the basis
 of positional correlations without the account of multiple
 tries~\cite{comparative} with (or, for the fit ``0'', without) correction
for GMF (lower values correspond to more probable sources of cosmic rays);
(5), distance to the object (in Megaparsecs, assuming $H=70$~km$\cdot {\rm
s}^{-1} \cdot {\rm Mpc}^{-1}$);
(6) and (7) --  object position
 in the J2000 equatorial coordinates (in degrees); (9), the probability
 to be randomly associated with a cluster's source with charge assignment
 (8) (lower values correspond to better fits); (10), ``yes'' means that the
 combination of distance and charge assignment requires new physics or
 cosmology. See text (and footnote on page~\pageref{footnote}) for
 details.} \lineup
\begin{flushright}
 {\footnotesize
\begin{tabular}{@{}cccccccccc}
\br
No. & Name & Class &$P_{\rm class}$ & $d$
& $\alpha $ &
 $\delta $ & Fit & $P$& new phys.\\
 (1) & (2)  & (3)   & (4)       & (5)       & (6) & (7) & (8) & (9) &
 (10) \\
 \mr
1& TXS 1055$+$567    &BLL &$4\cdot 10^{-4}$&617& 164.66 & 56.47 & N1 &
 0.21&yes\\
 2& 3EG~J1052$+$5718  &$\gamma $-s.&-&-& (163.2) & (57.3)
&\multicolumn{2}{c}{extended}   &-\\
3& NGC 3642          &Sy~\cite{Veron2003}  &0.62&23& 170.57 & 59.07 & N2 &
 0.05&no\\
 &                   &    &&&        &       & P  & 0.13&no\\
4&MCG $+$10-16-111   &Sy~\cite{Lipovetsky}  &0.013&120& 169.85 & 59.35 &
N2
 & 0.09&yes\\
 &                   &    &&&        &       & P  & 0.11&no\\
5& NGC 3488          &coll.&0.15&43&165.35 & 57.68 & P  & 0.23&no\\
6& NGC 3517          &coll.&0.27&118&166.40 & 56.52 & 0  & 0.54&yes\\
7& M 108             &IR~\cite{HCN}  &0.11&10& 167.87 & 55.67 & 0  &
 0.35&no\\
 &                   &    &&&        &       & Z  & 0.45&yes\\
8& NGC 3471          &IR~\cite{PDS}  &0.47&30& 164.79 & 61.53 & N2 &
 0.14&no\\
 9& IC 694            &IR,coll.&0.15&45& 172.13 & 58.57 & N2 & 0.04&no\\
\br
\end{tabular}
 }
\end{flushright}
\end{table}
\subsection{Notes on individual sources.}
\paragraph{TXS 1055$+$567.} This is an outstanding low-energy peaked BL Lac
type object listed as a confirmed ``BL'' in the V\'eron
catalog~\cite{Veron2003}. With its visual magnitude of
$15^{\rm\scriptscriptstyle m}.8$, strong dominance of the optical over
X-ray emission and possible association with a EGRET sub-GeV source, this
object satisfies all conditions for a BL Lac to be a potential cosmic-ray
source, as discussed in Refs.~\cite{TT:GMF,TT:BL,BL:EGRET,BL:index}. There
is a recent evidence, based on the high-precision HiRes stereo data, that
the objects of this particular subclass of BL Lac's may emit {\em neutral}
ultra-high-energy particles~\cite{BL:HiRes,BL:index}. This makes its
association with the cluster at the charge assignment N1 --- with a
neutral Yakutsk primary --- rather plausible, despite a comparatively low
quality of the fit ($\chi ^2/$d.o.f.$=4.72/8$; the BL Lac position
corresponds almost exactly to the highest-probability point).

\paragraph{3EG~J1052$+$5718.} This gamma-ray source~\cite{3EG} is probably
associated with the same BL Lac TXS 1055$+$567~\cite{3EG} (see also
Ref.~\cite{BL:EGRET}). The source is relatively weak and variable; its
gamma-ray location is determined with rather poor accuracy, and the $95\%$
C.L.\ position contours extend for several degrees. We thus do not quote
the probability values for this object, but plot instead the position
contours in Fig.~\ref{fig:skymap}. The connection of sub-GeV gamma-rays to
the emission and propagation of ultra-high-energy cosmic rays has been
pointed out several
times~\cite{BL:EGRET,Berezinsky,Blasi:gamma,Fargion:gamma}.

\paragraph{NGC~3642 and MCG $+$10-16-111} are nearby Seyfert galaxies listed,
correspondingly, in the catalogs~\cite{Veron2003} and \cite{Lipovetsky}
and satisfying the criteria of Ref.~\cite{Uryson} to be possible sources
of cosmic rays.

\paragraph{NGC~3488 and NGC~3517} are interacting pairs of galaxies from
the Vorontsov-Veliaminov catalog~\cite{VV}. Colliding galaxies were
considered as possible sources of ultra-high-energy cosmic rays in
Refs.~\cite{AGASA1999,colliding}.

\paragraph{NGC~3556 and NGC~3471} are luminuos infrared galaxies from the
HCN~\cite{HCN} and PDS~\cite{PDS} catalogs, correspondingly. Possible
origin of energetic cosmic particles in luminuos infrared and starburst
galaxies was discussed in Refs.~\cite{LIG}.

\paragraph{IC~694} is a colliding pair of galaxies~\cite{VV} with a
starburst region~\cite{PDS}. A possible relation of this object to the
AGASA triplet was discussed in Ref.~\cite{AGASA1999}. It enters the list
of luminuos infrared galaxies with a gamma-ray flux in excess of the
expected GLAST sensitivity~\cite{Torres-LIG}.

\paragraph{A note on NGC~3610 and NGC~3613.}
These two objects, located not far from the direction of the cluster,
entered the list of twelve dead quasar candidates of Ref.~\cite{Torres}
selected from the catalog~\cite{NOG} by five parameters. It was their
proximity to the AGASA triplet which was responsible to the correlations
of these objects with cosmic rays reported in Ref.~\cite{Torres} (two
sources and three cosmic rays contributed most of the observed
coincidences). However, updated galaxy catalogs~\cite{LEDA} quote fainter
corrected blue magnitude for these objects, so that they no longer satisfy
quite involved criteria of Ref.~\cite{Torres}; that is why we do not
include them in the list of possible candidates. We note in passing that
the list of objects selected from the catalog~\cite{LEDA} by criteria of
Refs.~\cite{Torres,NOG} no longer exhibits any correlations with energetic
cosmic rays~\cite{comparative}.

\subsection{A simultaneous interaction?}
One of possible ways to explain the existence of a tight cluster of
charged particles is to suppose that their {\em path} through the magnetic
fields was short enough. This might be the case if all five particles, or
some of them, originated in an interaction of a single particle which took
place not far from the Earth. This could be a photodesintegration of a
nuclei, or a Z-burst. In that case, the time delays between the arrivals
of particles should be approximately proportional to their angular
deflections from the direction to the place where the interaction
happened. This is hardly possible in our case because the particles
arriving later falled between the two which arrived first (see
Table~\ref{Tab:events}). This does not exclude, however, the scenario when
numerous enough particles were created in a single explosive event like a
gamma-ray burst and, being randomly deflected by magnetic fields, continue
to reach the Earth for very long time.

\section{Conclusions}
\label{sec:concl}
The existence of a cluster of five cosmic rays with energies exceeding
$4\cdot 10^{19}$~eV (AGASA scale), observed by three independent
experiments, supports the conjecture of clustering
of ultra-high-energy cosmic rays and of their astrophysical origin.
Three of the events did not enter the original AGASA sample, on
the basis of which the conjecture was formulated, thus allowing for the
first evidence for a triplet purely independent of the original claim. Our
study does not address the statistical significance of clustering but
suggests that the five cosmic rays were of common origin.

At high Galactic latitude corresponding to the cluster direction, the
Galactic magnetic field is weak and poorly known (the effect of
unknown halo fields may dominate over that of fading disk field). The
existence of a cluster of charged particles may be considered as an
argument towards the weak intergalactic magnetic field
scenario~\cite{IGMF:Dolag}.

The fits for the possible position of the source of these five cosmic
rays are better with the assumption of regular magnetic deflection than
without it, suggesting that at least some of the particles were charged.
Best-fit source positions are listed in Table~\ref{Tab:fits}, but the
uncertainties in the positions are large, as shown in
Fig.~\ref{fig:skymap}.

Within particular published models of the Galactic magnetic field, the fits
are not better than without magnetic deflections at all.
Better fits appear in a two-parametric analysis and correspond to shifts
in other directions. These facts may be interpreted as a signature for
unknown regular Galactic or extragalactic magnetic field component in this
direction.

Within given experimental accuracy, the fits cannot prefer, nor exclude,
any particular astrophysical source. Relative estimates of the quality of
fits for different sources may be found in Table~\ref{Tab:sources}. Under
particular charge assignments, the BL Lac type object TXS~1055$+$567, the
EGRET source 3EG~J1052$+$5718, probably associated with this BL Lac, and
the colliding starburst galaxy IC~694 may be related to the
clustered cosmic rays, but the evidence is far from being firm. Clearly,
better angular resolution (like that of the HiRes stereoscopic experiment)
is required for firm identification of the cosmic-ray sources from other
clusters. This is a necessary direction of further studies.

One should note that the physical scenarios discussed here are strongly
constrained by the propagation of energetic particles in cosmic background
radiation. Within the conventional physics and cosmology, the Z-burst
model is excluded~\cite{SemikozNeutrino}; the protons of the energies
considered here can propagate for about several hundred megaparsecs; the
nuclei photodisintegration scenario is relevant for the source distances
up to 100~Mpc~\cite{AGASA:nuclei}. As for the neutral particles, photons
may propagate for $5\dots 10$~Mpc while neutrons of this energy can travel
only $\sim 1$~Mpc before decay. Looking at the distance to the sources
(Col.~5 of Table~\ref{Tab:sources}), one may note that in some cases,
the explanation of the association of a given source with the cluster
within a scenario indicated in Col.~8 requires new, unconventional
physics or cosmology (see Col.~10 of the table).

Future investigations should include also attempts to identify the
chemical composition of the clustered primary particles both by
statistical methods and on the (less precise) case-by-case study of the
shower profiles, muon content etc.

Much better knowledge of the Galactic magnetic field, especially
at high Galactic latitudes, is required for identification of the sources
of charged ultra-high-energy cosmic-ray particles. On the other hand, once
these sources are identified, the cosmic-ray arrival directions may trace
the magnetic fields with some accuracy.

\ack
The author is indebted to the Yakutsk collaboration for sharing their
revised data prior to publication and for P.~Tinyakov who shared his code
for calculating deflections in the Galactic magnetic field in the model of
Ref.~\cite{TT:GMF}. I thank D.~Gorbunov, D.~Semikoz, P.~Tinyakov and
I.~Tkachev for numerous helpful discussions and comments on the
manuscript. I acknowledge interesting discussions with
K.~Belov, M.~Kachelriess, V.~Kuzmin, M.~Libanov, M.~Pravdin, V.~Rubakov,
S.~Sibiryakov, M.~Teshima and E.~Waxman.
This work was supported in part by
the INTAS grant 03-51-5112,  by the grant of the
President of the Russian Federation NS-2184.2003.2, by the grant of
the Russian Science Support Foundation and by the
fellowship of the "Dynasty" foundation (awarded by the Scientific
board of ICFPM).
This research has made use of the following online services: the VizieR
catalogue access tool, CDS, Strasbourg, France; the HyperLeda
database~\cite{LEDA};
the NASA/IPAC Extragalactic Database (NED)~\cite{NED};
the Vorontsov-Velyaminov catalog of interacting
galaxies~\cite{VV}.

\end{document}